\documentclass[epj]{svjour}

\usepackage{graphics}
\usepackage{amssymb}
\usepackage{cite}
\usepackage{graphicx,color}
\usepackage{soul}
\usepackage{amsmath}

\newcommand{\dd}{\mathrm{d}}
\newcommand{\td}[2]{\frac{\dd #1}{\dd #2}}
\newcommand{\pd}[2]{\frac{\partial #1}{\partial #2}}
\newcommand{\fd}[2]{\frac{\delta #1}{\delta #2}}
\newcommand{\mean}[1]{\langle #1 \rangle}
\newcommand{\Int}[1]{\int\dd #1\;}
\newcommand{\IInt}[3]{\int_{#2}^{#3}\dd #1\;}
\newcommand{\path}[1]{\int \mathcal{D}#1\;}

\renewcommand{\vec}[1]{\mathbf #1}

\newcommand{\al}{\alpha}

\newcommand{\sig}{\sigma}
\newcommand{\x}{\vec r}

\newcommand{\noist}{\boldsymbol\xi}
\newcommand{\noisr}{\boldsymbol\chi}
\newcommand{\noisc}{\boldsymbol\zeta}
\newcommand{\tr}{\tau_\text{r}}
\newcommand{\Dr}{D_\text{r}}
\newcommand{\tf}{\text{eff}}
\newcommand{\Dt}{D_\text{t}}
\newcommand{\Da}{D_\text{a}}
\newcommand{\rc}{r_\text{c}}
\newcommand{\rs}{r_\text{s}}
\newcommand{\ec}{\epsilon_\text{c}}

\newcommand{\id}{\mathbf 1}

\newcommand{\lp}{\ell_\text{p}}
\newcommand{\rhob}{\rho_\text{b}}

\begin{document}

\title{Applicability of Effective Pair Potentials for Active Brownian Particles}
\author{Markus Rein \and Thomas Speck}
\institute{Institut f\"ur Physik, Johannes Gutenberg-Universit\"at Mainz, Staudingerweg 7-9, 55128 Mainz, Germany}

\date{Received: date / Revised version: date}

\abstract{We have performed a case study investigating a recently proposed scheme to obtain an effective pair potential for active Brownian particles [Farage \emph{et al.}, Phys. Rev. E \textbf{91}, 042310 (2015)]. Applying this scheme to the Lennard-Jones potential, numerical simulations of active Brownian particles are compared to simulations of passive Brownian particles interacting by the effective pair potential. Analyzing the static pair correlations, our results indicate a limited range of activity parameters (speed and orientational correlation time) for which we obtain quantitative, or even qualitative, agreement. Moreover, we find a qualitatively different behavior for the virial pressure even for small propulsion speeds. Combining these findings we conclude that beyond linear response active particles exhibit genuine non-equilibrium properties that cannot be captured by effective pair interaction alone.
}

\maketitle


\section{Introduction}

Active Brownian particles belong to a larger class of models that are currently studied for their collective dynamical behavior~\cite{vics12}. It is a minimal model of diffusing, interacting spherical particles, each of which is propelled along an axis. This axis is not fixed but undergoes, in the simplest case free, rotational diffusion. What has stimulated attention is the observation of aggregation and \emph{dynamical clustering} resembling the liquid-vapor phase separation of passive particles, but caused by a dynamical instability. This phenomenon has been observed in experiments~\cite{theu12,pala13,butt13} and has been studied theoretically mostly for purely repulsive particles~\cite{yaou12,redn13,bial13,wyso14,cate15}, for which the clustering is a genuinely non-equilibrium phase transition requiring sufficiently strong driving. But even for particles with attractive interactions, which do undergo liquid-vapor separation in equilibrium, the self-propulsion strongly influences the non-equilibrium phase behavior~\cite{line12,redn13a,mogn13,mani15}.

Due to the directed motion time-reversal symmetry is broken and the active particles are driven into a non-equilibrium steady state implying that heat is constantly dissipated. However, there are no persistent global particle currents since the orientations decorrelate within a finite time. This has led to suggestions that active Brownian particles might be amenable to an effective thermodynamic description despite being strongly driven~\cite{taka15}. For example, the mentioned dynamical clustering can be described via an effective free energy for a coarse-grained density~\cite{cate13,cate15,spec15} and (in the absence of torques) the pressure allows for an equation of state~\cite{yang14,Brady_2014,wink15,solo15}. In contrast, effects like a negative interfacial tension~\cite{bial15} cannot be explained based on equilibrium concepts.

When only particle positions are observed the directed motion implies a memory and thus a non-Markovian process~\cite{bert13}. In the past there has been considerable interest how such non-Markovian processes can be approximated by a Markov process. Following these ideas, there have been two recent proposals for active Brownian particles discussing approximate schemes how to obtain an effective, \emph{equilibrium} Markov process. The first by Farage~\emph{et al.}~\cite{fara15} is based on Fox's approximation, originally derived for a single degree of freedom~\cite{fox86_func,fox86_uniform}. The second scheme by Maggi~\emph{et al.}~\cite{magg15,marc16a} employs the unified colored noise approximation~\cite{jung87,hang95}. For the latter also a connection to thermodynamics has been discussed applying equilibrium statistical mechanics to the effective pair potential~\cite{marc15}.

Both schemes consider the dilute limit of two particles interacting via a pair potential. Here we follow the route of Farage~\emph{et al.} and derive the effective pair potential for the Lennard-Jones potential. We are interested to which degree such an effective pair potential can be transfered to suspensions of many interacting particles. We discuss bounds imposed by the mapping and study numerically a passive suspension of particles interacting via the effective pair potential in two and three dimensions. All simulations are carried out at a single, moderately high density. We compare both the structure and the virial pressure of the active suspension with the mapped passive suspension. Based on the numerical results we argue that such a mapping becomes reliable only close to equilibrium, where it corresponds to a linear-response relation.


\section{Model and Theory}

\subsection{Model}

In this work, we study $N$ active Brownian particles moving in $n\in\{2,3\}$ spatial dimensions. Particles interact via the pair potential $u(r)$ with total potential energy $U(\{\vec r_i\}) = \sum_{i<j}u(|\vec r_i - \vec r_j|)$ and forces $\vec F_i = -\nabla_i U$, where $\nabla_i$ denotes the gradient with respect to the position $\vec r_i$ of particle $i$. The coupled equations of motion read for the positions
\begin{equation}
  \label{eq:lang}
  \dot\x_i  = v_0 \vec e_i + \mu_0 \vec F_i + \sqrt{2\Dt}\noist_i,
\end{equation}
and for the orientations (employing Stratonovich)
\begin{equation}
  \label{eq:orient}
  \dot{\vec{e}}_i = \sqrt{2\Dr} \noisr_i  \times \vec e_i.
\end{equation}
Here, $\mu_0$ is a (bare) mobility obeying the Einstein relation $\mu_0=\beta \Dt$, where $\beta=(k_\text{B}T)^{-1}$ is the inverse temperature. Particles move with constant speed $v_0$ due to the self-propulsion. The corresponding orientations $\vec e_i$ diffuse on the unit sphere (circle) with ($n-1$) degrees of freedom and a time correlation that decays exponentially with the time constant $\tr = [(n-1)\Dr]^{-1}$ defined by the rotational diffusion constant $\Dr$. The noise contributions $\boldsymbol\eta_i=\noist_i,\noisr_i$ are modeled as Gaussian white noise with zero mean and variance $\mean{\eta^\al_i(t) \eta^\beta_j(t')}=\delta^{\al\beta}\delta_{ij}\delta(t-t')$, where upper indices label vector components.

Specifically, we study ABPs interacting by either the Lennard-Jones (LJ) potential
\begin{equation}
  \label{eq:lj}
  u_{\text{LJ}}(r)=4\epsilon \left[\left(\frac{\sigma}{r}\right)^{12}-\left(\frac{\sigma}{r}\right)^6\right], 
  \quad \rc = \sqrt[6]{2}\sigma,
\end{equation}
with length scale $\sig$, or the purely repulsive Weeks-Chandler-Andersen (WCA) potential~\cite{weeks_role_1971}
\begin{equation}
  \label{eq:wca}
  u_{\text{WCA}}(r) =
  \begin{cases}
    u_{\text{LJ}}(r) + \epsilon\, & (r \leq \rc) \\
    0 & (r > \rc),
  \end{cases}
\end{equation}
which is derived from the LJ potential  by truncating the LJ potential in its minimum $\rc$ and shifting it by $\epsilon$ to ensure energy conservation.


\subsection{Effective pair potential}

The first step is the exact transformation of the Markov dynamics of Eqs.~\eqref{eq:lang} and~\eqref{eq:orient} into a non-Markovian process, which is obtained by integrating out the angular degrees of freedom~\cite{yaou12,fara15}. The resulting stochastic differential equations
\begin{equation}
  \label{eq:coarse_dyn}
  \dot\x_i  = \noisc_i + \mu_0 \vec F_i + \sqrt{2\Dt}\vec \noist_i
\end{equation}
describe the non-Markovian time-evolution of the particle positions with the self-propulsion and rotational dynamics now included in the stochastic process $\noisc_i$ with $|\noisc_i|=v_0$, which has zero mean and variance
\begin{equation}
  \label{eq:corr}
  \mean{\zeta_i^\al(t)\zeta_j^\beta(t')} = 
  \frac{v_0^2}{n}\delta^{\al\beta}\delta_{ij} e^{-|t-t'|/\tr}.
\end{equation}
The directed motion thus introduces an exponential memory with correlation time $\tr$. In the limit $\tr\to0$ we recover a Markov process for the particle positions alone, where orientations diffuse so fast that the system effectively behaves like an equilibrium system at the elevated temperature $T_\tf=T(1+\Da/\Dt)$ with active contribution $\Da=v_0^2\tr/n$ to the diffusion coefficient.

We now follow Farage~\emph{et al.} to approximate the non-Markovian process by a Markov process~\cite{fara15}. This will involve a number of uncontrolled approximations. First, we have to relax the normalization condition and assume that $\noisc_i$ is described by a Gaussian process (see Refs.~\citenum{marc16a,fodo16} for simulations of the resulting model). In appendix~\ref{sec:fox}, we outline the derivation for many degrees of freedom. In appendix~\ref{sec:two} we derive the effective force between two particles in the limit of a dilute suspension
\begin{equation}
  \label{eq:force_eff}
  \vec F^\tf(r) = \frac{1}{D(r)}\left[\Dt\vec F(r) - \beta^{-1}\nabla D(r)\right]
\end{equation}
as a function of particle separation $r$. Here, $\vec F(r)=-\nabla u(r)$ is the passive force and
\begin{equation}
  \label{eq:diff_eff}
  D(r) = \Dt + \frac{\Da}{1+2\mu_0\tr\Delta_r u}
\end{equation}
is an effective diffusion coefficient.\footnote{Note that we find a factor $2$ in front of $\tr$, which is missing in Ref.~\citenum{fara15}. This has been commented previously by Cates and Nardini~\cite{cate16}.} The radial Laplacian in $n$ dimensions reads $\Delta_ru=u''+(n-1)u'/r$. Hence, $\vec F^\tf$ is defined by the two-body force, the number of spatial dimensions $n$, and the self-propulsion through the speed $v_0$ and the correlation time $\tr$. Moreover, the force points along the particle separation $\vec e_r$, $\vec F^\tf(r)=F^\tf(r)\vec e_r$. We can thus always integrate the effective force to obtain an effective potential $u^\tf(r)$ so that $\vec F^\tf=-\nabla u^\tf$.

It is instructive to expand the effective force to linear order of the correlation time $\tr$. With
\begin{equation}
  D(r) \approx \Dt + \Da(1-2\mu_0\tr\Delta_r u)
\end{equation}
we obtain
\begin{multline}
  \label{eq:force_eff_lin}
  F^\tf = -\frac{u'}{\hat D} + \hat\tr\left[u'''+(n-1)\frac{u''}{r}-(n-1)\frac{u'}{r^2}\right] \\ - \frac{\hat\tr}{\hat D}\beta\left[u'u''+(n-1)\frac{(u')^2}{r}\right],
\end{multline}
where
\begin{equation}
  \hat D = \frac{\Dt+\Da}{\Dt} = \frac{T_\tf}{T}, \qquad
  \hat\tr = \frac{2\Da\tr}{\hat D}.
\end{equation}
The integral of the last term in Eq.~\eqref{eq:force_eff_lin} is not a  closed expression, and is absent from the first-order expansion of the result found by Maggi~\emph{et al.}~\cite{magg15},
\begin{equation}
  \label{eq:ueff}
  u^\tf = \frac{u}{\hat D} + \frac{\hat\tr}{2\hat D}\beta(u')^2 - \hat\tr u'' - (n-1)\hat\tr\frac{u'}{r}.
\end{equation}
Hence, ignoring this term, both methods (almost) agree for appropriately redefined diffusion coefficient $\hat D$ and (small) correlation time $\hat\tr$.

\subsection{Persistence length}

With $\sig$ the length scale of the potential, the small dimensionless parameter required in the derivation of the effective force Eq.~\eqref{eq:force_eff} reads
\begin{equation}
  \label{eq:bound}
  \tau = \frac{\tr}{\sigma^2/\Dt} = \frac{1}{n-1}\frac{\Dt}{\Dr\sig^2} \ll 1
\end{equation}
independent of propulsion speed $v_0$. The directed motion can also be characterized by the ``persistence length'' $\lp=v_0\tr$, which quantifies the typical length over which particles remember their orientations. The agreement of the effective potential Eq.~\eqref{eq:ueff} with Ref.~\citenum{magg15} suggests that actually this persistence length is the small parameter since to lowest order
\begin{equation}
  \label{eq:pers}
  \frac{\hat\tr}{\sig^2} \approx \frac{2}{n}\left(\frac{v_0\tr}{\sig}\right)^2 
  \ll 1,
\end{equation}
which now does depend on the propulsion speed.

From now on we will employ dimensionless quantities so that lengths are expressed in units of $\sigma$, times in units of the translational Brownian time $\sigma^2/\Dt$, and energies in units of the thermal energy $\beta^{-1}$. In these units $v_0$ corresponds to the translational P\'{e}clet number and $\tr=\tau$. For colloidal spherical particles, the rotational diffusion coefficient is not a free parameter but set by the boundary condition. Assuming no-slip, it follows as $\Dr=3\Dt/\sig^2$ with $\tau=[3(n-1)]^{-1}$. For the dynamical clustering of repulsive particles, at least a persistence length of $\lp\sim15$ is required (cf. phase diagrams in Refs.~\citenum{redn13a,spec15}), which is beyond the range for which we expect the mapping to be useful. Instead, in the following we will treat $\tau$ as a free parameter using $\tau=0.025$ ($\tau=0.05$) with speeds $v_0\leq40$ corresponding to $\lp\leq1$ ($\lp\leq2$).

\subsection{Admissible potential strengths}

\begin{figure}
  \centering	
  \includegraphics{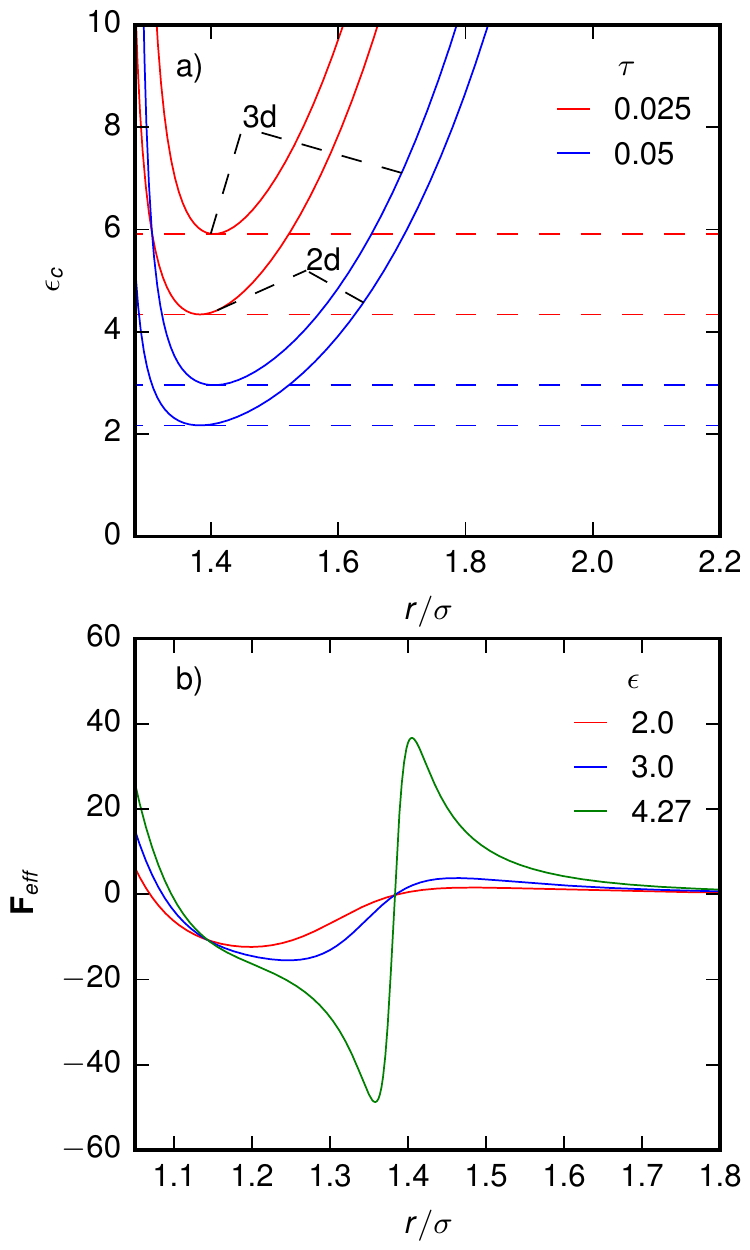}
  \caption{(a)~Critical values of $\ec$ for the two values of $\tau$ as a function of separation. The dashed lines indicate the minimal value $\epsilon_0$. Choosing $\epsilon>\epsilon_0$ leads to singularities in the effective force. (b)~Effective force of the LJ potential in two dimensions for $\tau=0.025$ and $v_0=20$ varying $\epsilon$. When approaching the singularity ($\epsilon\approx4.3$), the effective force develops a deep minimum implying highly attractive forces.}
  \label{fig:map_eps_crit}
\end{figure}

\begin{figure*}[t]
  \centering
  \includegraphics{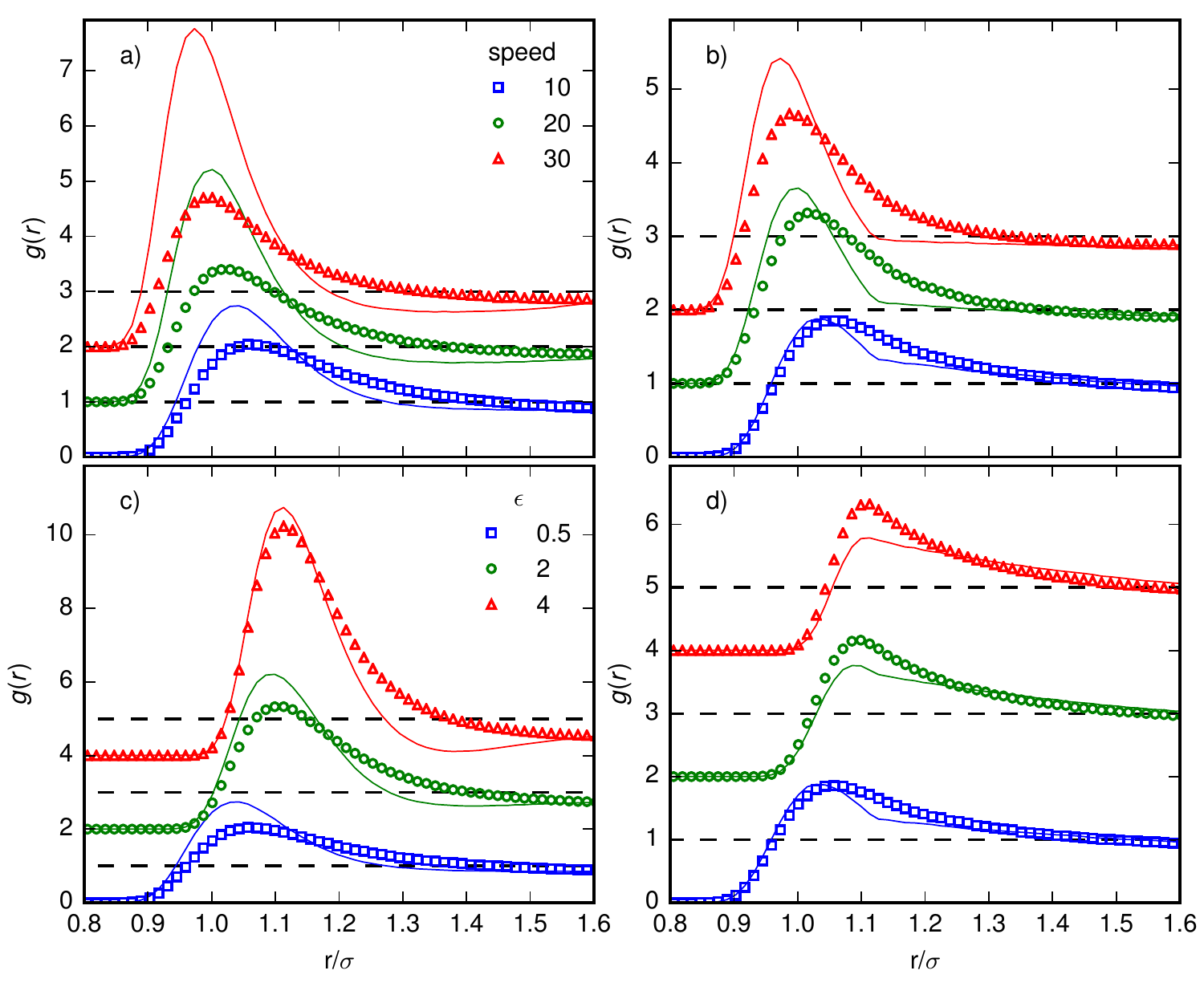}
  \caption{Comparison of the pair distribution functions $g(r)$ in two dimensions: Active Brownian particles (symbols) and the mapped passive system (solid lines) at $\tau=0.025$ and area fraction $\phi = 0.3$. The first row discusses the role of speed at $\epsilon = 0.5$, for (a)~the LJ system and (b)~the WCA system. In the second row $\epsilon$ is varied at constant speed $v_0=10$ for (c)~the LJ system and (d)~the WCA system. Vertical black dashed lines indicate $g(r)=1$. Results are shifted vertically for clarity.}
  \label{fig:v_eps_eff}
\end{figure*}

For the LJ potential Eq.~\eqref{eq:lj} it is straightforward to determine the effective force $F^\tf=-(u'+D')/D$ although the resulting expression is somewhat cumbersome. For distances $r<\rc$ the effective forces for the LJ and the WCA potential agree, with the effective force for the WCA potential becoming discontinuous at $r=\rc$. In Ref.~\citenum{fara15} the limit Eq.~\eqref{eq:bound} is mentioned but it is implied that, through including higher order of $\tau$, this mapping could produce useful results even beyond the limit Eq.~\eqref{eq:bound}. From Eq.~\eqref{eq:diff_eff} it becomes clear that nevertheless the condition
\begin{equation}
  \label{eq:domain}
  1 + 2\tau\Delta_ru(r) > 0
\end{equation}
has to hold for the effective diffusion $D(r)$ to be positive and finite, and thus for the mapping to be at least possible. This condition restricts the range of admissible $\epsilon$. For the LJ potential in three dimensions, we start by noticing that $\Delta_ru(r)>0$ for $r<\rs=\sqrt[6]{22/5}\simeq1.28$ and changes its sign for $r>\rs$. Hence, it follows that $1+2\tau\Delta_ru(r)\geq 1$ for $r\leq \rs$, in agreement with Eq.~\eqref{eq:domain}. For $r>\rs$ we find that the admissible potential strength $\epsilon<\epsilon_0$ is bounded by the minimum of
\begin{equation}
  \label{eq:sing_eps3}
  \ec(r) = \frac{r^{14}}{240\tau(r^6 - \rs^6)},
\end{equation}
which, for $r>\rs$, is a positive, convex function with a minimum value of $\epsilon_0\simeq0.15/\tau$ at position $r_0=\sqrt[6]{77/10}\simeq1.41$, cf. Fig.~\ref{fig:map_eps_crit}(a). It does not depend on the propulsion speed $v_0$ and decreases with increasing $\tau$. As a consequence, in the range $r>\rs$ for every $\epsilon\geq\epsilon_0$ a singularity in the effective force occurs at some distance $r$, where $\rs<r\leq r_0$. Since $\rs>\rc=\sqrt[6]{2}\simeq1.12$, the effective force arising from the WCA potential is always defined, regardless of the chosen values for $\tau$ and $\epsilon$.

Fig.~\ref{fig:map_eps_crit}(b) shows the effective force as a function of distance $r$ for several values of $\epsilon$ and $\tau=0.025$. It shows an attractive domain for $r<r_0$ and a repulsive domain for $r>r_0$. Approaching the critical value of $\epsilon_0$, the minimum of the effective force decreases, implying highly attractive forces. By further increasing $\epsilon$, a steep minimum develops in the attractive domain, which rises to a steep maximum in the repulsive domain. This demonstrates that for admissible but relatively large values of $\epsilon$ the effective force shows a strong dependence on the fixed length $r_0$ independent of the self-propulsion parameters.

For the LJ potential in two dimensions, the singularity shifts to $\rs=\sqrt[6]{4}\simeq1.26$. The critical $\epsilon$ becomes
\begin{equation}
  \label{eq:sing_eps2}
  \ec(r) = \frac{r^{14}}{288\tau(r^6 - \rs^6)}
\end{equation}
and position and value of the minimum shift to $r_0=\sqrt[6]{7}\simeq 1.38$ and $\epsilon_0\simeq 0.11/\tau$, cf. Fig.~\ref{fig:map_eps_crit}(a).

\section{Numerical results}
\label{sec:numerics}

\subsection{Structure in two dimensions}

We now investigate to which extent the effective pair potentials can be transfered to many-body suspensions and reproduce the structure of Brownian active particles at moderate densities. To this end we perform Brownian dynamics simulations for both the active Brownian particles and the mapped effective passive system in $n=2$ dimensions. We use $N=5000$ particles and employ a quadratic box with periodic boundary conditions. The simulation of active Brownian particles is divided in three steps. Using a timestep of $\delta t=10^{-5}$, a system of passive particles is equilibrated for 10, afterward active Brownian particles are relaxed into the steady state for 50 before we measure the quantities of interest over another 100 Brownian times. For the mapped system, passive particles interacting by the effective potential are simulated over 100 Brownian times. For the LJ potential we employ a cutoff of $2$. Packing fractions presented in this work are defined by $\phi=\rhob\pi/(2n)$ with density $\rhob=N/L^n$ and are not rescaled employing an effective particle diameter.

\begin{figure}[t]
  \centering
  \includegraphics{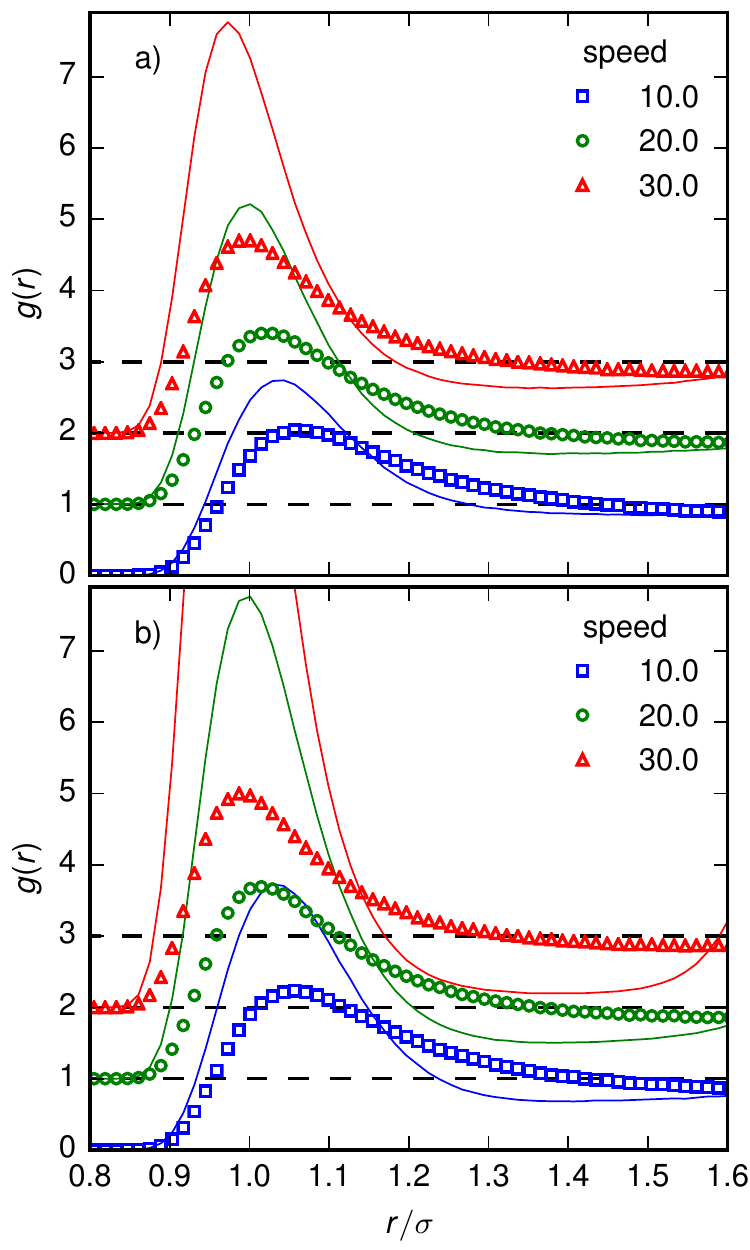}		
  \caption{Comparison of the pair distribution function for the LJ potential at $\epsilon = 0.5$ showing active Brownian particles (symbols) and the mapped passive system (solid lines). (a)~$\tau=0.025$ and (b)~$\tau=0.05$ for several speeds $v_0$. Simulations are carried out at area fraction $\phi=0.3$.}
  \label{fig:tau_eff}
\end{figure}

We first focus on the structure through the pair correlation function $g(r)$. With increasing speed the position of the first peak of the pair correlation function shifts towards smaller distances and increases in height, compare Fig.~\ref{fig:v_eps_eff}(a) for the LJ potential and (b) for the WCA potential. Higher speeds increase the probability of observing smaller particle separations, since particles are able to more easily enter the repulsive domain of the interaction force. While the effective potential qualitatively reproduces this finding, deviations between the active and the effective descriptions arise for both potentials, which become more severe as $v_0$ becomes larger. For the LJ potential, the pair correlations stemming from the passive simulations exceed the active result until approximately $\rc$. For larger separations, $g(r)$ in the effective description falls below the active one. As a consequence, particle separations are shifted towards smaller values. For the WCA potential, cf. Fig.~\ref{fig:v_eps_eff}(b), the effective $g(r)$ also steepens towards smaller separations with increasing speed. In contrast to the LJ potential, the relative difference between passive and active simulations remains smaller. Furthermore, due to the discontinuity of the effective force at the cutoff $\rc$ of the WCA potential, the pair correlations resulting from the effective simulations are not smooth at $\rc$. The discontinuity of the effective LJ force has a smaller impact on $g(r)$ due to the larger cutoff.

Increasing $\epsilon$ for the LJ potential at fixed particle speed $v_0=10$, the first peak of the pair correlation function increases in height while its position shifts to larger particle separations. The pair correlations resulting from the effective description qualitatively reproduce this finding, cf. Fig.~\ref{fig:v_eps_eff}(c), while now the agreement of peak height and position improves as we increase $\epsilon$. The first peak rising from the WCA potential, cf. Fig.~\ref{fig:v_eps_eff}(d), also shifts towards larger separations, but does not increase as strongly as for the LJ potential due to the lack of attractive forces in the active case. Again, the pair correlations for the mapped system shifts towards larger separations, while the increment of the central peak is now too small. We stress that the speed $v_0=10$ is significantly smaller than the critical speed for dynamical clustering to occur at the area fraction of $\phi=0.3$.

In Fig.~\ref{fig:tau_eff} we show the effect of changing the reorientation time $\tau$. While the validity of the effective potential is restricted by the product $\epsilon \tau$, $\tau$ and $\epsilon$ represent different influences on the dynamics of active systems. Whereas both parameters influence the phase behavior of the suspension, $\epsilon$ scales the interaction strength of particles, and $\tau$ defines their reorientation time. Accordingly, the dependence of the deviations between effective and active description in the pair correlation function is more severe for $\tau$ than for $\epsilon$. Doubling the reorientation time by increasing $\tau$ from 0.025, Fig.~\ref{fig:tau_eff}(a) to 0.05, Fig.~\ref{fig:tau_eff}(b), at fixed $\epsilon = 0.5$, the deviations of the main peak of $g(r)$ increase even more strongly. Again, the critical values $\ec(\tau=0.025)\approx4.3$, and $\ec(\tau=0.05)\approx2.2$ causing a singularity in the effective diffusivity are far from being reached.

\begin{figure}[b!]
  \centering
  \includegraphics{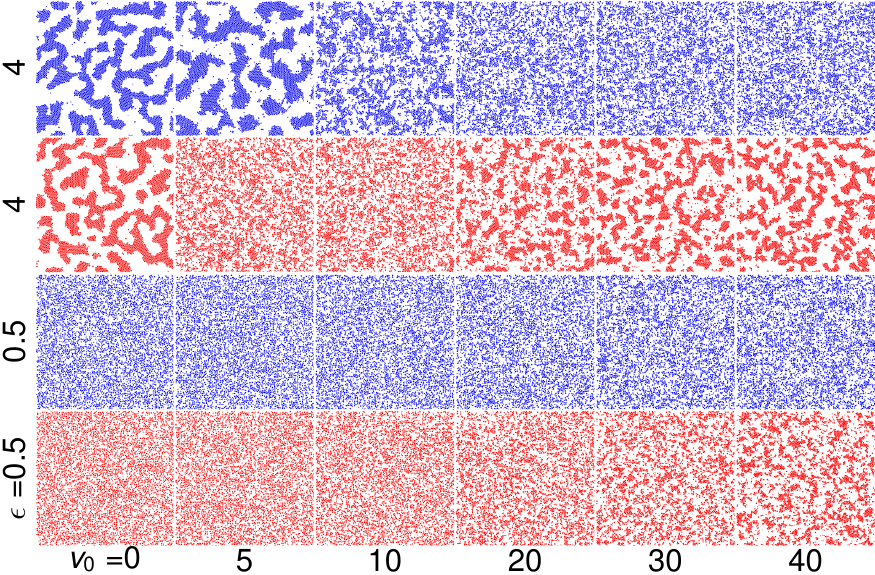}		
  \caption{Simulation snapshots for the LJ potential. We compare active Brownian particles (blue) to the effective, mapped system (red) at $\tau=0.025$. Particle positions are plotted at $t=140$ ($t=90$) for the active (mapped) system.}
  \label{fig:snapsphase}
\end{figure}

Finally, in Fig.~\ref{fig:snapsphase} snapshots of the active and mapped system for $\tau=0.025$ are presented as a function of interaction strength $\epsilon$ and speed $v_0$. Shown are the particles positions after $t=140$ ($t=90$) for the active (mapped) system. Starting with $\epsilon=0.5$, both the active and mapped system yield a homogeneous phase agreeing with the small deviations of the pair correlation function in this range. For the largest speeds the mapped system shows the formation of domains, which is absent in the original active system. Increasing $\epsilon$ to 4, liquid-vapor phase separation and the formation of dense domains is observed for $v_0=0$. At higher speeds both systems return to the homogeneous state in line with Ref.~\citenum{redn13a}. While the mapped system qualitatively reproduces the transition to the homogeneous phase, it reenters a phase-separated state for speeds $v_0\gtrsim20$. In contrast, the active system remains in the homogeneous state and only enters the dynamical cluster phase at much higher speeds beyond what we simulate here.

\subsection{Virial pressure}

In thermal equilibrium structure and thermodynamics are tied to each other. Proposing a mapping to an effective equilibrium system thus begs the question to which degree this connection is valid. Accordingly, beyond the structure of the pair correlations for the active and mapped systems, a second interesting quantity is the scalar virial pressure. This quantity has been studied quite intensely recently. It has not been considered in Ref.~\citenum{fara15}. Following Refs.~\citenum{Brady_2014,wink15}, for the active system it is determined via \begin{equation}
  p(v_0, \phi, \tau) = \frac{\rhob}{nN}\left[ v_0\sum_i \mean{\vec e_i \cdot \vec r_i} +  \sum_{i<j} \mean{\x_{ij} \cdot \vec F_{ij}} \right],
\end{equation}
where $\mean{\cdot}$ denotes the average in the steady state. The first term accounts for the pressure contribution resulting from the active forces employing the absolute coordinates $\x_i$ taking into account crossings of the periodic boundaries. It implicitly depends on $\tau$ through performing the average since the reorientation time influences the temporal evolution of the orientations $\vec e_i$. The second term accounts for the conservative pair interactions. Accordingly, $\sum_{i<j}$ is a double sum including all particle pairs and $\x_{ij}=\x_i-\x_j$ describes the vector connecting particle pairs with pair forces $\vec F_{ij}=\vec F(|\x_{ij}|)$. In the mapped system, particles are passive and the active forces are effectively included in the interaction potential, leading to the conventional virial pressure
\begin{equation}
  p^\tf(v_0,\phi,\tau) = \frac{\rhob}{nN}\sum_{i<j} \mean{\vec r_{ij} \cdot \vec F^\tf_{ij}}
\end{equation}
of the mapped system, where $\vec F^\tf_{ij}=\vec F^\tf(|\x_{ij}|)$ are the effective pair forces Eq.~\eqref{eq:force_eff}.

\begin{figure}[t]
  \centering
  \includegraphics{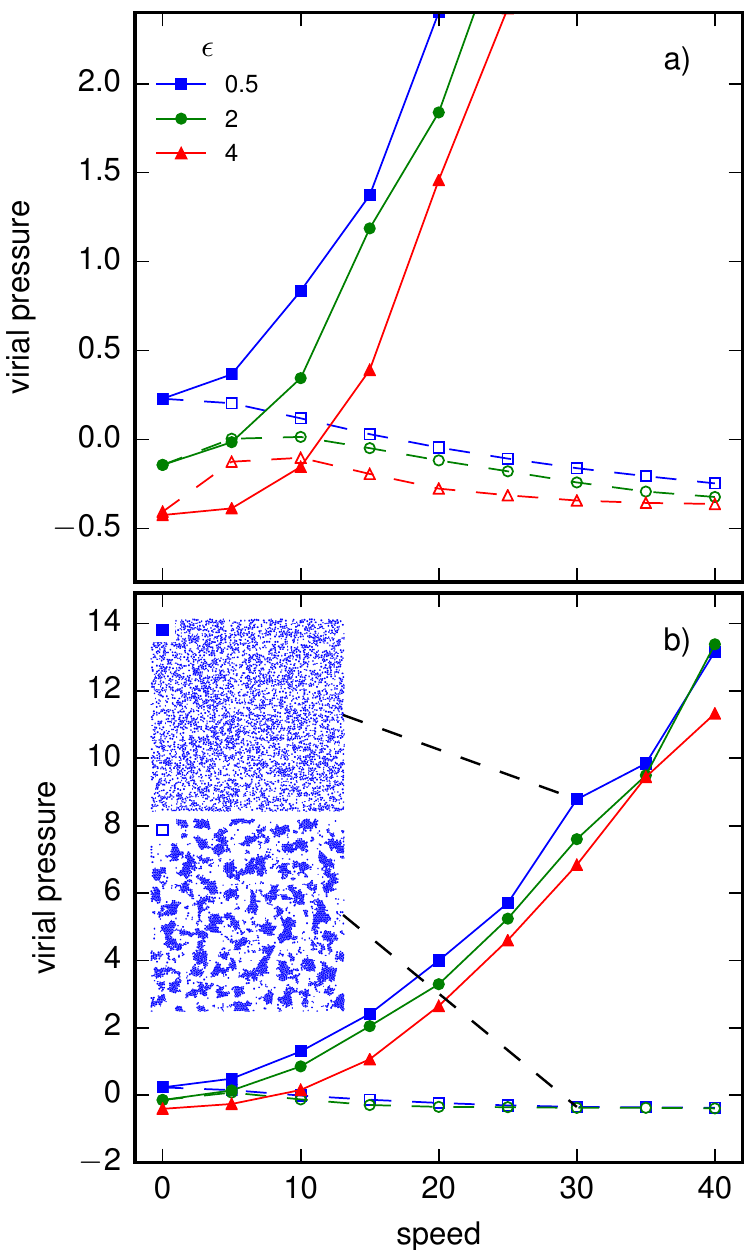}
  \caption{Virial pressure \emph{vs.} speed for three different potential strengths $\epsilon$ comparing active Brownian particles (solid symbols) to the effective, passive system (open symbols) interacting by the LJ potential for (a)~$\tau=0.025$ and (b)~$\tau=0.05$. Insets in (b) show snapshots after $t=140$ ($t=90$) for the active (mapped) system for the same speed $v_0=30$, indicating the qualitative difference of structures. In (b), $\epsilon=4$ for $\tau=0.05$ cannot be mapped anymore and only the active suspension is shown. Simulations are carried out at area fraction $\phi=0.3$.}
  \label{fig:pres_eff_lj}
\end{figure}

In Fig.~\ref{fig:pres_eff_lj} we compare the virial pressures measured for the active and the mapped system for the LJ potential. For $\tau=0.025$ [Fig.~\ref{fig:pres_eff_lj}(a)] and $\epsilon=0.5$ both systems remain in the homogenous state. The virial pressure of the active system increases strongly with speed. In contrast, the virial pressure for the mapped system slightly decreases since the increasing speed implies larger attractions in the effective pair potential. For larger $\epsilon=4$, active particles show a phase-separated state for small speeds, in which the pressure increases only slightly. In this regime the pressure of the mapped system rises before decreasing again for large speeds. For $\tau=0.05$ [Fig.~\ref{fig:pres_eff_lj}(b)] we find qualitatively the same discrepancies.

\subsection{Three dimensions}

\begin{figure}[t]
  \centering
  \includegraphics{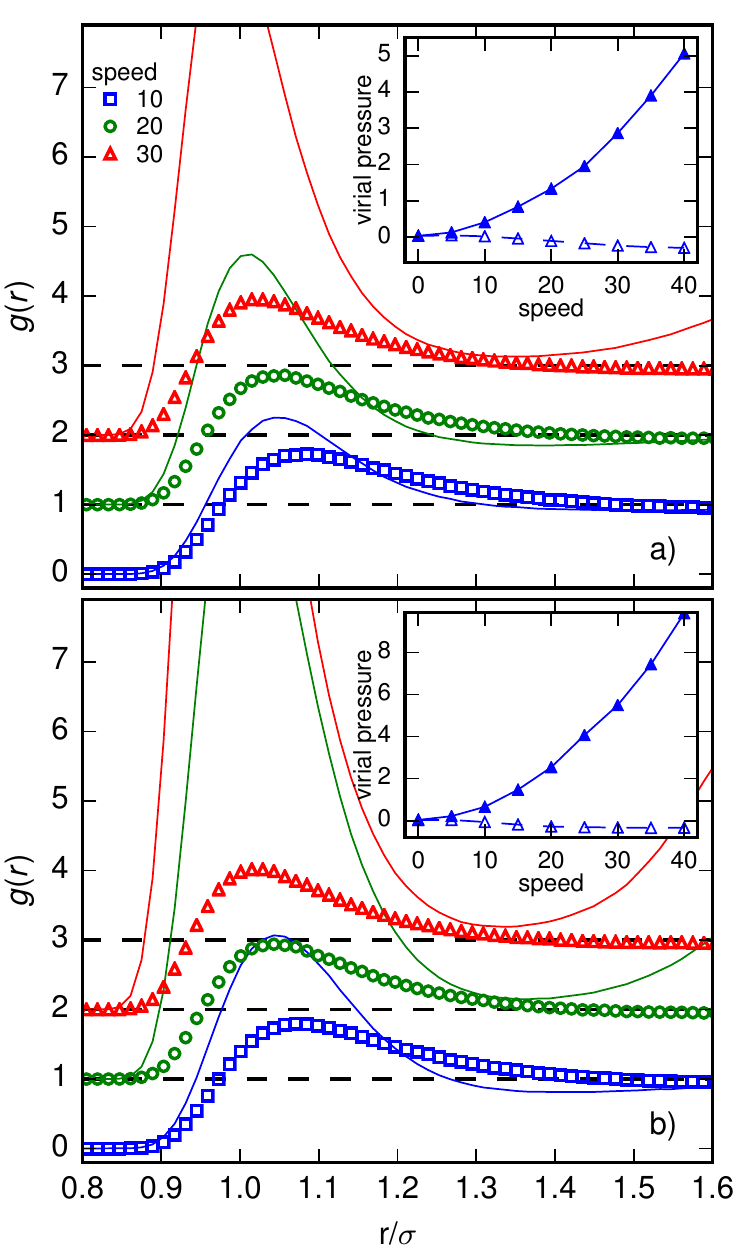}
  \caption{Comparison of the pair correlation function for active and effective mapped systems in $n=3$ spatial dimensions. Simulations are carried out at $\phi=0.1571$ and $\epsilon=0.5$ with (a)~$\tau=0.025$ and (b)~$\tau=0.05$. Insets show the corresponding virial pressure for different speeds and identical parameters.}
  \label{fig:3d}
\end{figure}

As done in the previous section, simulations of $N=5000$ active Brownian particles are compared to simulations of passive particles interacting by the effective potential but now for three spatial dimensions. We again use a cubic simulation box of length $L$ employing periodic boundary conditions. In Fig.~\ref{fig:3d}, the pair correlation function as well as the virial pressure is shown for two values of $\tau$ varying the speed. The area fraction is lowered to $\phi = 0.1571$ to account for particle distances comparable to the two dimensional case. Comparing Fig.~\ref{fig:tau_eff} to Fig.~\ref{fig:3d}, the first central peak of the structure factor in the active system slightly increases in height and shifts to smaller separations for higher speeds. This is also reproduced by the mapped, passive system, as observed in the two dimensional case. Furthermore, the position of the central peak in the pair correlation of the effective system is at slightly smaller separations compared to the active case and rises to a larger maximum of $g(r)$. The deviations of the active and the mapped system in the pair correlation increase with speed, in line with the results of the two dimensional system. For the active system remaining in the homogenous state, the virial pressure again increases strongly with speed as in the two dimensional system for both $\tau=0.025$ and $\tau=0.05$. Also in three dimensions we find that for $\epsilon=2,4$ the passive systems form domains at higher speeds $v_0$.

\section{Discussion}

We have compared the dynamics of active Brownian particles interacting via a LJ (or WCA) potential to a mapped system of passive Brownian particles interacting by an effective pair potential following the route proposed by Farage \emph{et al.}~\cite{fara15}. The mapped description requires the speed $v_0$, the reduced reorientation time $\tau$, the spatial dimensions $n$, and the interaction potential of active Brownian particles as input. For small activity parameters $\tau$ and $v_0$ the state of active particles is described qualitatively correct by the mapped system. In particular, the directed motion induces (additional) attractions in the mapped potential capturing the tendency of active particles to accumulate. However, the mapping tends to overestimate these attractions, leading to a much stronger first peak in the pair distribution function than observed for the active system. This discrepancy becomes more severe as propulsion speed $v_0$ and correlation time $\tau$ are increased, and less severe as the potential depth $\epsilon$ is increased (as long as it is sufficiently small compared to the value $\epsilon_0$ at which the singularity in the pair potential appears). The mapped system qualitatively reproduces entering the homogeneous phase as $v_0$ is increased but due to the overestimated attractions the mapped system again orders at higher speeds $v_0\gtrsim20$ while the active suspension remains homogeneous.

The shape of the effective pair potential includes short-ranged repulsion, attractions, and long-ranged repulsion introducing a second length scale. From this shape one expects an equilibrium cluster phase with stable domains of a typical size set by this second length scale~\cite{scio04}. Somewhat surprisingly (and presumably unphysical), the second length scale $r_0$ only depends on dimensions and not on the propulsion parameters. It is presently not clear whether the observed clusters~\cite{redn13a,mani15} in the active suspension correspond to the final steady state or a long-lived metastable state. The equilibrium state of the LJ potential is a single domain, which would require another transition if the steady state of the active suspension should indeed be a cluster phase.

We have also studied the virial pressure of active and mapped passive suspensions. Here the discrepancy is even more drastic showing a qualitative different nature. The pressure in the active suspensions rises strongly as $v_0$ is increased. In contrast, due to the stronger attractions, the pressure in the mapped passive suspensions drops to slightly negative values.

The main conclusion from these observations is that an effective mapping is applicable close to equilibrium for the structure (not the pressure). We finally note that in the linear response regime an effective potential is always possible as can be seen as follows: Denoting $\psi(\{\x_i\};\lp)$ the steady state distribution of the active system parametrized by the persistence length $\lp$, the average of any observable $A(\{\x_i\})$, in particular the pair distribution function $g(r)$, can be written
\begin{equation}
  \mean{A} = \mean{A}_0 + \lp\mean{AB}_0 + \mathcal O(\lp^2)
\end{equation}
with another observable $B=\partial\ln\psi/\partial\lp|_{\lp=0}$. This observable can be absorbed into an effective potential $U^\tf=U-\lp B$, which to first order produces the same average $\mean{A}$ as in the active system. Neglecting three-body and higher interactions then leads to a pair potential like Eq.~\eqref{eq:ueff} with the precise coefficients determined by the mapping procedure.

\section{Conclusions}

In a numerical case study, we have investigated to which extent effective isotropic pair potentials can be used to describe active Brownian particles. For the Lennard-Jones potential, we have shown that the product $\epsilon\tau$ (independent of the propulsion speed $v_0$) restricts the formal applicability of the mapping. Within the admissible range, from numerical simulations we find that the structural agreement deteriorates quickly for both larger speeds $v_0$ and larger correlation times $\tr$. For the WCA potential there is no formal restriction but numerically we find the same trend for the structure. Notwithstanding the issue of neglected three-body and higher interactions, we conclude that the mapping can capture small perturbations in the linear response regime with the reduced persistence length $\lp/\sig=v_0\tr/\sig$ as the small parameter. For the virial pressure we already find differences in the linear response regime (\emph{i.e.}, different slopes for small $v_0$). This points to the importance of genuine non-equilibrium effects that cannot be captured by effective attractions while neglecting dissipation.

\section*{Acknowledgments}

We thank Jonathan T. Siebert for support to verify the active Brownian particle simulations, and ZDV Mainz for the computing time on MOGON. We acknowledge financial support by the DFG through the priority program SPP 1726 (grant number SP 1382/3-1).

\section*{Author contributions}

M.R. and T.S. designed the research and jointly wrote the manuscript, M.R. wrote the code, performed simulations, and analyzed the data.


\appendix

\section{Fox's approximation}
\label{sec:fox}

Fox's result~\cite{fox86_func} for a single degree of freedom $\dot x=\mu_0F+\zeta$ with noise correlations
\begin{equation*}
  C(t-t') = \mean{\zeta(t)\zeta(t')} = \frac{\Da}{\tr}e^{-|t-t'|/\tr}
\end{equation*}
states that the non-Markovian process is approximated by a Markov process, the probability $\psi(x,t)$ of which evolves according to
\begin{equation*}
  \partial_t\psi = -\partial_x(\mu_0 F\psi) + 
  \Da\partial_x^2(G\psi), \quad
  G = \frac{1}{1-\mu_0\tr F'},
\end{equation*}
where the derivative of the force arises from the chain rule.

For the generalization to many degrees of freedom $x=\{x_k\}$ we reexamine the route followed by Farage~\emph{et al.}~\cite{fara15}. Making the distinction between stochastic variables $\xi$ and their values $x$, the equations of motion are $\dot\xi_k=\mu_0F_k+\zeta_k$ with noise correlations $C_{kl}(t)=\delta_{kl}C(t)$, cf. Eq.~\eqref{eq:corr}. The evolution equation for the joint probability
\begin{equation*}
  \psi(x,t) = \mean{\delta(x-\xi(t))}
\end{equation*}
reads
\begin{equation*}
  \partial_t\psi = -\sum_k\partial_k(\mu_0F_k\psi) - \sum_k\partial_k\mean{\delta(x-\xi(t))\zeta_k},
\end{equation*}
where we denote
\begin{equation*}
  \mean{f} = \path{\zeta} P[\zeta] f[\zeta]
\end{equation*}
the path integral over the noise history with Gaussian weight $P[\zeta]$. We now use the identity
\begin{equation*}
  P[\zeta]\zeta_k = -\Int{s} C(t-s)\fd{P}{\zeta_k(s)}
\end{equation*}
and functional integration by parts to obtain
\begin{multline}
  \label{eq:app}
  \mean{\delta(x-\xi(t))\zeta_k} = \Int{s}C(t-s)\left\langle \fd{[\delta(x-\xi(t)]}{\zeta_k(s)} \right\rangle \\
  = -\sum_l\partial_l\Int{s}C(t-s) \left\langle \delta(x-\xi(t))\fd{\xi_l(t)}{\zeta_k(s)} \right\rangle.
\end{multline}
We define two matrices with components
\begin{equation*}
  A_{lk}(t,s) = \fd{\xi_l(t)}{\zeta_k(s)}, \qquad
  J_{ln}(t) = \left.\pd{F_l}{x_n}\right|_{\xi(t)}
\end{equation*}
leading to the differential equation
\begin{equation*}
  \td{}{t}A_{lk} = \fd{\dot\xi_l(t)}{\zeta_k(s)} = \mu_0\sum_nJ_{ln}A_{nk} + \delta_{lk}\delta(t-s)
\end{equation*}
with solution (for $t>s$)
\begin{equation*}
  \vec A(t,s) = \exp\left\{ \mu_0\IInt{s'}{s}{t}\vec J(s') \right\} 
  \approx e^{\mu_0(t-s)\vec J(t)},
\end{equation*}
which is a matrix exponential. With the final approximation we can pull the matrix $\vec A$ out of the brackets in Eq.~\eqref{eq:app} with the symmetric Hessian $\vec J$ evaluated at $x$. The $\delta$-function then contracts to the joint probability $\psi(x,t)$.

The final step is to evaluate the integral (extending the upper limit to infinity)
\begin{equation}
  \label{eq:psi}
  \IInt{s}{0}{\infty} C(s)e^{\mu_0s\vec J} = \Da(\id-\mu_0\tr\vec J)^{-1} = \Da\vec G(x),
\end{equation}
which is obtained by expanding the matrix exponential in a power series and then resumming terms requiring that the matrix norm $||\mu_0\tr\vec J||\leq1$ is bounded. The result for the approximate Markov process is then
\begin{equation*}
  \partial_t\psi = -\sum_k\partial_k(\mu_0F_k\psi) + \Da \sum_{k,l}\partial_k\partial_l(G_{kl}\psi).
\end{equation*}
Quite in contrast to Ref.~\citenum{fara15}, we find that the effective diffusion matrix has off-diagonal terms mediated by the forces. To obtain a diagonal form, one possible (though rather unjustified) operation that preserves the determinant of $\vec G$ is
\begin{equation}
  \label{eq:G}
  \vec G \approx \frac{\id}{\det(\id-\mu_0\tr\vec J)} 
  \approx \frac{\id}{1-\mu_0\tr\text{Tr}(\vec J)},
\end{equation}
where in the last step we have again appealed to the smallness of the correlation time $\tr$.

\section{Two interacting particles}
\label{sec:two}

We now consider two particles at positions $\x_1$ and $\x_2$ with separation $\x=\x_1-\x_2$ and $r=|\x|$. The pair potential $u(r)$ is isotropic and thus all quantities depend only on $\x$, in particular we have $\vec F_1=\vec F$ and $\vec F_2=-\vec F$, and $\nabla_1=\nabla$ and $\nabla_2=-\nabla$. Taking into account the translation noise, the evolution of the pair distribution Eq.~\eqref{eq:psi} with Eq.~\eqref{eq:G} becomes
\begin{equation*}
  \partial_t\psi = -2\nabla\cdot(\mu_0\vec F-\Dt\nabla)\psi + 
  2\Da\nabla^2\left[\frac{1}{1-\mu_0\tr\text{Tr}(\vec J)}\psi\right].
\end{equation*}
For the trace of the $(2n\times2n)$ Hessian we find
\begin{equation*}
  \text{Tr}(\vec J) = \sum_{k=1}^{2n} \partial_kF_k 
  = \nabla_1\cdot\vec F_1+\nabla_2\cdot\vec F_2
  = 2\nabla\cdot\vec F.
\end{equation*}
Rearranging terms leads to
\begin{equation*}
  \partial_t\psi = -2\nabla\cdot D\left[\beta\vec F^\tf - \nabla\right]\psi
\end{equation*}
with effective force $\vec F^\tf(r)$ and diffusion coefficient $D(r)$ given in the main text in Eqs.~\eqref{eq:force_eff} and~\eqref{eq:diff_eff}.


\end{document}